\documentclass[showpacs,preprintnumbers,amsmath,amssymb,prb]{revtex4}

\def\XXint#1#2#3{{\setbox0=\hbox{$#1{#2#3}{\int}$}
     \vcenter{\hbox{$#2#3$}}\kern-.5\wd0}}

\usepackage{graphicx}
\usepackage{subfigure}
\usepackage{bm}

\begin{document}

\title{Cascades of phase transitions in spiral magnets caused by dipolar forces}

\author{O.\ I.\ Utesov$^1$}
\email{utiosov@gmail.com}
\author{A.\ V.\ Syromyatnikov$^{1,2}$}
\email{asyromyatnikov@yandex.ru}

\affiliation{$^1$National Research Center "Kurchatov Institute" B.P.\ Konstantinov Petersburg Nuclear Physics Institute, Gatchina 188300, Russia}
\affiliation{$^2$St.\ Petersburg State University, 7/9 Universitetskaya nab., St.\ Petersburg, 199034
Russia}

\date{\today}

\begin{abstract}

We present a mean-field theory describing the influence of long-range dipolar forces on the temperature transition from the paramagnetic to ordered phases in frustrated Heisenberg spiral magnets. It is shown that the dipolar interaction produces a cascade of first- and second- order phase transitions between the paramagnetic and the spiral states upon temperature decreasing. Depending on system parameters, the following intermediate phases can arise: an incommensurate and a commensurate sinusoidally modulated states, spiral phases in which perpendicular spin components have different amplitudes and are modulated with the same and with different wave vectors. We distinguish six possible sequences of phase transitions upon temperature decreasing at least four of which were observed before experimentally in specific compounds. It is found that the action of dipolar forces cannot always be modeled even qualitatively by small one-ion anisotropic spin interactions. We demonstrate that the dipolar interaction is responsible for successive phase transitions in the triangular-lattice multiferroic MnI$_2$: almost all available experimental findings are described quantitatively within the mean-field theory by taking into account the exchange, the dipolar and small symmetry-allowed anisotropic spin interactions.

\end{abstract}


\maketitle

\section{Introduction}

Frustration can have a dramatic impact on properties of magnetic systems leading to novel phenomena which have being extensively studied in recent years: various spin-liquid phases, novel phase transitions, and order-by-disorder phenomena, to mention just a few. \cite{balents} In particular, frustration changes the type of transitions to magnetically ordered phases in Heisenberg antiferromagnets (HAFs) on a (stacked) triangular lattice and in frustrated HAFs with a spiral magnetic ordering. The order parameter acquires additional symmetry elements that leads to changing the type of the phase transition in three-dimensional (3D) systems (the continuous transition in non-frustrated magnets vs.\ the first-order one in frustrated systems), to a novel pseudo-universal behavior in 3D $XY$ systems, and to the stabilization of a chiral spin-liquid phase upon cooling before the onset of Berezinskii-Kosterlitz-Thouless transition in 2D systems. \cite{kawa,sasha}

Weak low-symmetry spin interactions, which are always present in real materials, complicate further the behavior of frustrated systems upon temperature decreasing. They can lead, for example, to a crossover to another critical behavior near the critical point, to a changing the type of the phase transition, and to a splitting of the phase transition into a sequence of different phase transitions. In particular, it is well known that dipolar forces, which are always present in real compounds, lead to the splitting of the transition to the ordered state with $120^\circ$ magnetic structure into three successive transitions in $XY$ HAFs on the stacked triangular lattice. \cite{shiba,gekht} Three successive transitions take place upon the temperature decreasing: the second-order transition from the paramagnetic (PM) phase to an incommensurate sinusoidally-modulated (ICS) state, the second-order transition to an incommensurate phase in which two components of magnetic moments are modulated with different wave vectors and have different amplitudes (an elliptic phase), and, finally, the first-order transition occurs to the commensurate phase with the conventional $120^\circ$ magnetic structure. The difference between temperatures of these three transitions is governed by the ratio of the characteristic dipolar energy $\omega_0$ and the exchange coupling constant $J$ which is usually small in real materials. However three successive phase transitions with these two incommensurate intermediate phases were really observed in particular triangular $XY$ HAFs (e.g., in $\rm RbFeCl_3$) with $J\sim\omega_0\sim 1$~K (see Refs.~\cite{shiba,gekht}).

Frustrated Heisenberg magnets in which the spiral magnetic ordering arises due to the competition between different exchange interactions fall into the same universality classes as triangular HAFs. \cite{kawa} To the best of our knowledge, the impact of the dipolar interaction on transitions to magnetically ordered phases has not been discussed yet in such models. On the other hand, such investigation would be of particular interest due to the great attention devoted in recent years to multiferroics with spiral magnetic orderings appearing due to frustrated exchange interactions. \cite{nagaosa} This attention is stimulated by a possible application of such compounds in the spin-related electronics. Multiferroics $\rm MnI_2$ (Refs.~\cite{nagaosa,sato,cable,mni1,mni2,mni3}) and $\rm MnWO_4$ (Refs.~\cite{nagaosa,mnw1,mnw2,mnw3}) are promising candidates for such analysis because their exchange coupling constants are small ($\lesssim 1$~K). Besides, the magnetocrystalline anisotropy is expected to be very small because Mn$^{2+}$ ions are in spherically symmetric states with the orbital and the spin moments $L=0$ and $S=5/2$, respectively. Then, the dominating low-symmetry interaction in these compounds is the dipolar one. It was found experimentally that these materials show the cascade of phase transitions upon temperature decreasing with the ICS and elliptical intermediate phases.

\begin{figure}
  \noindent
  \includegraphics[scale=1]{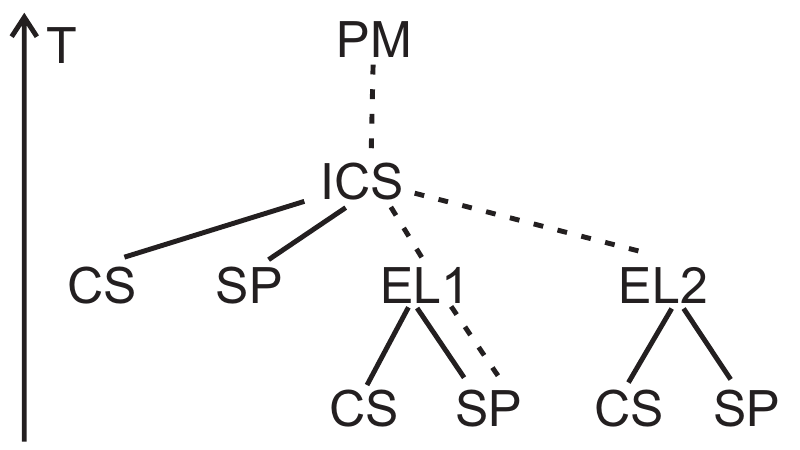}
  \caption{Possible sequences of temperature phase transitions in frustrated spiral Heisenberg antiferromagnets with dipolar forces. PM, ICS, CS and SP stand for the paramagnetic, incommensurate sinusoidally modulated, commensurate sinusoidally modulated and spiral phases, correspondingly. EL1 and EL2 are elliptical phases in which perpendicular spin components have different amplitudes and are modulated with the same (EL1) and with different (EL2) wave vectors. Transitions of the first and of the second order are shown by solid and by dashed lines, respectively. The transition from the EL1 phase to the SP one can be either of the first or of the second order depending on the model parameters (see the text).
  \label{scen}
	}
\end{figure}

We develop a mean-field theory in Sec.~\ref{gen} describing frustrated spiral HAFs (including HAFs on the triangular lattice) with dipolar forces near the transition from the PM phase. Phases which can arise in this model are described: the ICS phase, the commensurate and the incommensurate spiral states, elliptical phases in which two components of the order parameter are modulated with the same and with different vectors. Six possible sequences of transitions to these phases are established which are summarized in Fig.~\ref{scen}. Phase transitions in MnBr$_2$, $\rm MnWO_4$, and in $XY$ HAFs on the stacked triangular lattice follow one of these six scenarios. It is shown that the transition from the PM state takes place to the ICS phase. Then, we extend in Sec.~\ref{gen} available theories devoted solely to the role of the dipolar interaction in MnBr$_2$ \cite{mnbr2} and in triangular $XY$ HAFs \cite{shiba}.

It is always tempting to model the action of dipolar forces by some short-range anisotropic spin interactions in theoretical considerations due to slow convergence of dipolar sums that requires using the inconvenient special resummation technique. We consider in Sec.~\ref{anisdisc} the possibility to reproduce all six scenarios obtained in Sec.~\ref{gen} by replacing the dipolar interaction (which is a source of a biaxial anisotropy in a system) by a short-range biaxial spin anisotropy. We find that only three scenarios can be reproduced by the single-ion anisotropy whereas all six scenarios arise in the case of the exchange biaxial anisotropy.

We describe quantitatively phase transitions in MnI$_2$ in Sec.~\ref{mni2} within the mean-field approach. It is shown that dipolar forces are indispensable for a proper description of available experimental data \cite{sato,cable}, but small symmetry-allowed easy axis and hexagonal anisotropies should be also taken into account. Besides, our analysis shows that Dzyaloshinskii-Moriya interaction (DMI) should arise in the spiral phase which is responsible for ferroelectric properties in this phase. The latter result is in accordance with recent experimental findings. \cite{mni3} MnI$_2$ follows one of six scenarios described in Sec.~\ref{gen} which is somewhat complicated by the small anisotropic interactions.

We present a summary of the results and our conclusion in Sec.~\ref{conc}. The mean-field expansion of the free energy and DMI in the spiral ferroelectric phase of MnI$_2$ are discussed in appendixes.

\section{Phase transitions in spiral Heisenberg magnets with dipolar forces}
\label{gen}

In this section, we discuss how dipolar forces change the transition from the PM to the ordered phase in frustrated spiral HAFs. We assume that the order parameter is small and develop the mean-field (Landau) theory. The corresponding mean-field energy reads as
\begin{equation}
\label{ham11}
  \mathcal{E} = \frac12\sum_{i,j} \left(J_{ij} \mathbf{s}_i \cdot \mathbf{s}_j + D^{\alpha \beta}_{ij} s^\alpha_i s^\beta_j\right),
\end{equation}
where the first and the second terms describe the exchange and the dipolar interactions, respectively, ${\bf s}_i$ is a mean magnetic moment which depends on $T$ and which is always smaller than the spin value $S$, and the summation is implied over repeated Greek letters which denote Cartesian components $x,y,z$. The dipolar tensor in Eq.~\eqref{ham11} has the form
\begin{equation}
  D^{\alpha \beta}_{ij} = \omega_0 \frac{v_0}{4\pi}
	\left( \frac{1}{R^3_{ij}} - \frac{3 R^\alpha_{ij} R^\beta_{ij}}{R^5_{ij}}\right),
\end{equation}
where $v_0$ is the unit cell volume and
\begin{equation}
  \omega_0=4\pi\frac{(g \mu_B)^2}{v_0}
\end{equation}
is the characteristic dipolar energy.

Within the mean-field approach, we obtain by expanding the free energy $\mathcal{F}$ up to the fourth order in $s$ (see Appendix~\ref{mf})
\begin{equation}
\label{f1}
  \mathcal{F}=\mathcal{E} + A T \sum_i s^2_i + B T \sum_i s^4_i,
\end{equation}
where $A$ and $B$ are values depending on $S$ only which are given by Eqs.~\eqref{a} and \eqref{b}, respectively. Introducing the Fourier transform
$
  \mathbf{s}_i=\frac{1}{\sqrt{N}} \sum_\mathbf{q} \mathbf{s}_\mathbf{q} e^{i\mathbf{q} \mathbf{R}_i},
$
we rewrite energy \eqref{ham11} near the transition from the PM phase as
\begin{equation}
\label{ham21}
   \mathcal{E} = \sum_{\mathbf{q}} \left( J_\mathbf{q} \delta_{\alpha \beta} + \frac12 D^{\alpha \beta}_\mathbf{q}\right) s^\alpha_\mathbf{q} s^\beta_{-\mathbf{q}}=\sum_{\mathbf{q}} \mathcal{H}^{\alpha\beta}_{\mathbf{q}} s^\alpha_\mathbf{q} s^\beta_{-\mathbf{q}},
\end{equation}
where $J_\mathbf{q} = \sum_{j\neq0} J_{0j} e^{i \mathbf{q}\mathbf{R}_j}$ and
$
  D^{\alpha \beta}_\mathbf{q} = \sum_{j\neq0} D^{\alpha \beta}_{0j} e^{i \mathbf{q}\mathbf{R}_j}
$.
Slowly convergent lattice sums in the latter expression are calculated below numerically by rewriting the sums in fast convergent forms (see Ref.~\cite{cohen}).

Tensor $\mathcal{H}^{\alpha\beta}_{\mathbf{q}}$ has three generally different eigenvalues $\lambda_{1,2,3}({\mathbf{q}})$ which are functions of $\bf q$. As it is seen from Eqs.~\eqref{f1} and \eqref{ham21}, the smallest eigenvalue $\lambda_1({\mathbf{q}}={\bf q}_{sin})$ and the corresponding eigenvector determine the free energy and the spin ordering in the ordered phase near the critical temperature. We consider below a typical situation of different minimum values of $\lambda_{1,2,3}({\mathbf{q}})$ (assuming that the smallest and the largest eigenvalues are $\lambda_1$ and $\lambda_3$, respectively) and an incommensurate value of ${\bf q}_{sin}$. Notice that ${\bf q}_{sin}\approx {\bf Q}$ at small dipolar interaction, where $\bf q= Q$ minimizes $J_{\bf q}$. The second-order transition from the PM phase to the ordered one takes place within the mean-field theory at a temperature $T_{N1}$ at which the bilinear term in the free energy changes the sign. Then, one obtains from Eqs.~\eqref{f1} and \eqref{ham21}
\begin{equation}
\label{tn1}
	T_{N1} = -\frac{\lambda_1({\mathbf{q}_{sin}})}{A}.
\end{equation}
The spin texture near $T=T_{N1}$ is determined by the eigenvector corresponding to $\lambda_1({\mathbf{q}}={\bf q}_{sin})$ which gives an incommensurate sinusoidally-modulated structure
\begin{equation}
\label{sdw}
  \mathbf{s}_i = \mathbf{a}_1 \sin \mathbf{q}\mathbf{R}_i + \mathbf{a}_2 \cos \mathbf{q}\mathbf{R}_i
\end{equation}
with $\mathbf{a}_1||\mathbf{a}_2$, $|{\bf a}_{1,2}|\propto s$, and $\mathbf{q}={\bf q}_{sin}$. Minimization of the free energy gives for its value and for the order parameter in the ICS phase
\begin{eqnarray}
\label{Fsin}
  \mathcal{F}_{ics} &=& -\frac{(\lambda_1({\mathbf{q}}_{sin})+AT)^2}{6BT}
	= -\frac{A^2(T_{N1}-T)^2}{6BT},\\
\label{s}
	s &=& \sqrt{\frac{2A}{3B}\frac{T_{N1}-T}{T}},
\end{eqnarray}
where Eq.~\eqref{tn1} is taken into account.

The model behavior at $T<T_{N1}$ depends strongly on values of its parameters. Let us consider possible ordered phases which can arise at $T<T_{N1}$. The first-order transition can happen from the ICS phase to that with the spiral order which is described by Eq.~\eqref{sdw} with $|\mathbf{a}_1|=|\mathbf{a}_2|$, $\mathbf{a}_1\perp\mathbf{a}_2$, and ${\mathbf{q}} ={\mathbf{q}}_{sp}$. The free energy of this state (denoted below as SP phase) and the transition temperature read as
\begin{eqnarray}
\label{Fsp}
  \mathcal{F}_{sp} &=& -\frac{([\lambda_1({\mathbf{q}}_{sp})+\lambda_2({\mathbf{q}}_{sp})]/2+AT)^2}{4BT},\\
\label{Tsp}
  T_{sp} &=& T_{N1}-\left( 1 +\sqrt{\frac{2}{3}}\right) S(S+1)(\lambda_1({\mathbf{q}}_{sp})+\lambda_2({\mathbf{q}}_{sp})-2\lambda_1({\mathbf{q}}_{sin})),
\end{eqnarray}
where $\lambda_1({\mathbf{q}})+\lambda_2({\mathbf{q}})$ reaches its minimum at ${\mathbf{q}} ={\mathbf{q}}_{sp}$ which is smaller than the minimum of $\lambda_1({\mathbf{q}})+\lambda_3({\mathbf{q}})$. Directions of $\mathbf{a}_1$ and $\mathbf{a}_2$ are determined by eigenvectors corresponding to $\lambda_1({\mathbf{q}}_{sp})$ and $\lambda_2({\mathbf{q}}_{sp})$.

It might happen that a commensurate vector $\mathbf{q}_{cs}$ lies not far from $\mathbf{q}_{sin}$ such that $2\mathbf{q}_{cs}$ or $4\mathbf{q}_{cs}$ are equal to a reciprocal lattice vector. Although $\lambda_1({\mathbf{q}})$ does not reach a minimum at $\mathbf{q}=\mathbf{q}_{cs}$, the free energy of the sinusoidally-modulated commensurate structures (CS) with $\mathbf{q}=\mathbf{q}_{cs}$ can become lower at some $T<T_{N1}$ than that of the ICS state. It can happen because summations over the lattice give different results at ${\bf q} = \mathbf{q}_{cs}$ and at an incommensurate $\bf q$ after substitution of Eq.~\eqref{sdw} to Eqs.~\eqref{ham11} and \eqref{f1}. Thus, one obtains for the free energy in this case
\begin{equation}
\label{Fcs}
  \mathcal{F}_{cs}=-\frac{(\lambda_1({\mathbf{q}}_{cs})+AT)^2}{4BT}.
\end{equation}
Notice the smaller numerical factor in the denominator of Eq.~\eqref{Fcs} as compared to that in Eq.~\eqref{Fsin} which makes possible the considered first-order transition from the ICS structure at the critical temperature
\begin{equation}
\label{Tcs}
  T_{cs}=T_{N1}-2 \left( 1 +\sqrt{\frac{2}{3}}\right) S(S+1)(\lambda_1({\mathbf{q}}_{cs})-\lambda_1({\mathbf{q}}_{sin})).
\end{equation}

A second-order transition can take place from the ICS to an elliptic structure described by Eq.~\eqref{sdw} with $|\mathbf{a}_1|\ne|\mathbf{a}_2|$, $\mathbf{a}_1\perp\mathbf{a}_2$, and $\mathbf{q}={\bf q}_{sin}$. Henceforth, it is called EL1 phase. One finds for the free energy of this state and the transition temperature
\begin{eqnarray}
\label{Fel}
  \mathcal{F}_{el1} &=& -\frac{3(\lambda_1({\mathbf{q}}_{sin})+AT)^2-2(\lambda_1({\mathbf{q}}_{sin})+AT)(\lambda_2({\mathbf{q}}_{sin})+AT) + 3(\lambda_2({\mathbf{q}}_{sin})+AT)^2}{16BT},\\
\label{Te}
  T_{el1} &=& T_{N1}-S(S+1)(\lambda_2({\mathbf{q}}_{sin})-\lambda_1({\mathbf{q}}_{sin})).
\end{eqnarray}

An elliptical structure in which two orthogonal spin components have different modulation vectors can arise also via a second-order transition from the ICS phase:
\begin{equation}
\label{el2order}
  \mathbf{s}_i = \mathbf{a}_1 \sin \mathbf{q}_{sin}\mathbf{R}_i + \mathbf{a}_2 \cos \mathbf{q}_2\mathbf{R}_i,
\end{equation}
where $|\mathbf{a}_1|\ne|\mathbf{a}_2|$, $\mathbf{a}_1\perp\mathbf{a}_2$, and ${\bf q}_2\ne \mathbf{q}_{sin}$. Vector $\mathbf{q}_{2}$ corresponds to the smallest eigenvalue of $\mathcal{H}^{\alpha\beta}_{\mathbf{q}}$ among eigenvectors which are perpendicular to the spin polarization in the ICS phase. Henceforth, this state is called EL2 phase. The free energy and the transition temperature read in this case
\begin{eqnarray}
\label{Fel2}
    \mathcal{F}_{el2} &=& -\frac{3(\lambda_1({\mathbf{q}}_{sin})+AT)^2-4(\lambda_1({\mathbf{q}}_{sin})+AT)(\lambda_1({\mathbf{q}}_{2})+AT) + 3(\lambda_1({\mathbf{q}}_{2})+AT)^2}{10BT},\\
\label{Te2}
  T_{el2} &=& T_{N1}-2S(S+1)(\lambda_1({\mathbf{q}}_{2})-\lambda_1({\mathbf{q}}_{sin})).
\end{eqnarray}

At $T\ll T_{N1}$, CS and SP phases are stable if $\lambda_1(\mathbf{q}_{cs}) < (\lambda_1(\mathbf{q}_{sp})+\lambda_2(\mathbf{q}_{sp}))/2$ and $\lambda_1(\mathbf{q}_{cs}) > (\lambda_1(\mathbf{q}_{sp})+\lambda_2(\mathbf{q}_{sp}))/2$, respectively. These conditions are  equivalent to $T_{cs}>T_{sp}$ and $T_{cs}<T_{sp}$, correspondingly. Conditions for transitions from the ICS to other phases mentioned above can be formulated in terms of inequalities between values of $T_{cs}$, $T_{sp}$, $T_{el1}$, and $T_{el2}$ given by Eqs.~\eqref{Tsp}, \eqref{Tcs}, \eqref{Te}, and \eqref{Te2}. The following six different scenarios can be distinguished which are schematically shown in Fig.~\ref{scen}.

(i) $T_{cs} > T_{sp},T_{el1},T_{el2}$. There is a first-order transition from the ICS to the CS state. The sequence of the phase transitions under temperature decreasing is the following: PM $\rightarrow$ ICS $\rightarrow$ CS. Phase transitions in MnBr$_2$ follow this scenario. \cite{mnbr2}

(ii) $T_{sp} > T_{cs},T_{el1},T_{el2}$. It is possible only if $\mathbf{q}_{sp} \neq \mathbf{q}_{sin}$. Then, there is a first-order transition from the ICS to the SP phase. The corresponding sequence is PM $\rightarrow$ ICS $\rightarrow$ SP. This scenario appears in MnI$_2$ which is complicated by small anisotropic spin interactions leading to an additional transition splitting the ICS state into two different ICS phases (see below).

(iii) $T_{el1} > T_{cs},T_{sp},T_{el2}$ and $T_{cs} > T_{sp}$. There is a second-order transition from the ICS to the EL1 structure and a first-order transition from the EL1 to the CS order. The corresponding sequence is PM $\rightarrow$ ICS $\rightarrow$ EL1 $\rightarrow$ CS. This succession of phase transitions was experimentally observed in MnWO$_4$. \cite{lauten}

(iv) $T_{el1} > T_{cs},T_{sp},T_{el2}$ and $T_{sp} > T_{cs}$. There is a second-order transition from the ICS to the EL1 structure. The subsequent transition from the EL1 to the SP phase is of the first-order type if $\mathbf{q}_{sp} \neq \mathbf{q}_{sin}$ and of the second-order type if $\mathbf{q}_{sp} = \mathbf{q}_{sin}$. The corresponding sequence is PM $\rightarrow$ ICS $\rightarrow$ EL1 $\rightarrow$ SP.

(v) $T_{el2} > T_{cs},T_{sp},T_{el1}$ and $T_{cs} > T_{sp}$. There is a second-order transition from the ICS to the EL2 structure and a first-order transition from the EL2 to the CS phase. The corresponding sequence is PM $\rightarrow$ ICS $\rightarrow$ EL2 $\rightarrow$ CS.

(vi) $T_{el2} > T_{cs},T_{sp},T_{el1}$ and $T_{sp} > T_{cs}$. There is a second-order transition from the ICS to the EL2 structure and a first-order transition from the EL2 phase to the spiral order. The corresponding sequence is PM $\rightarrow$ ICS $\rightarrow$ EL2 $\rightarrow$ SP. This scenario is realized in $XY$ HAFs on the stacked triangular lattice. \cite{shiba,gekht}

Notice that some fine details can be omitted in the picture just described. For instance, a small third harmonic of the modulation vector $\bf q$ can arise in Eq.~\eqref{sdw} which leads to a weak temperature dependence of $\bf q$ in the ICS state as it was observed \cite{mnbr2} in $\rm MnBr_2$. However we believe that apart from such fine details the above picture reflects all the possible phases and phase transitions which can arise in the considered model. Notice also that small anisotropic short-range spin interactions can complicate the above scenarios as it is demonstrated below by the example of MnI$_2$.

\section{Short-range anisotropic spin interactions}
\label{anisdisc}

In this section, we discuss the possibility to describe at least qualitatively the influence of the long-range dipolar interaction by some short-range spin interactions. We show first that although dipolar forces act as a source of low-symmetry biaxial anisotropy in a system, six scenarios of phase transitions discussed in Sec.~\ref{gen} cannot be reproduced by the one-ion biaxial anisotropy of the form
\begin{equation}
\label{biax}
  \mathcal{E}_{an}=\sum_i \left( E \left[ (s^x_i)^2 - (s^y_i)^2\right] - G (s^z_i)^2 \right).
\end{equation}
Let us assume for definiteness that $z$ is the easy axis and $x$ is the hard one:
\begin{equation}\label{bl}
  G>E>0, \qquad \delta {\cal A}=G - E.
\end{equation}
Particular analysis shows that the EL2 structure is always less energetically favorable than the EL1 state. Then, only relations between eigenvalues at ${\bf q}=\mathbf{q}_{sin}$ and the lowest eigenvalue among commensurate points $\lambda_1({\mathbf{q}}_{cs})$ determine the sequence of phase transitions. By the energy reason, the modulation vector in the SP and in the EL1 phases should be equal to $\mathbf{q}_{sin}$. As a result, the system can follow three different scenarios.

(i) A ``strong anisotropy'' scenario is realized when $T_{cs}>T_{el1}$ (see Eqs.~\eqref{Tcs} and \eqref{Te}) that reads as
\begin{equation}
 \lambda_2({\mathbf{q}}_{sin})-\lambda_1({\mathbf{q}}_{sin})=\delta {\cal A} > 2 \left( 1 +\sqrt{\frac{2}{3}}\right)(\lambda_1({\mathbf{q}}_{cs})-\lambda_1({\mathbf{q}}_{sin}))
\end{equation}
(notice that $T_{el1}$ is always larger than $T_{sp}$ given by Eq.~\eqref{Tsp} in the considered model with biaxial anisotropy \eqref{biax}). Thus, scenario (i) described in Sec.~\ref{gen} is realized.

(ii) A ``moderate anisotropy'' case implies
\begin{equation}
  2 (\lambda_1({\mathbf{q}}_{cs})-\lambda_1({\mathbf{q}}_{sin})) < \delta {\cal A} < 2 \left( 1 +\sqrt{\frac{2}{3}}\right)(\lambda_1({\mathbf{q}}_{cs})-\lambda_1({\mathbf{q}}_{sin}))
\end{equation}
that leads to the scenario (iii) described in Sec.~\ref{gen}. Thus, the phase diagram very similar to that of $\rm MnWO_4$ is obtained recently theoretically in Ref.~\cite{zh} in a spin model containing single-ion anisotropy \eqref{biax} and not containing the dipolar interaction.

(iii) A ``weak anisotropy'' case implies that
\begin{equation}
	\delta {\cal A} < 2 (\lambda_1({\mathbf{q}}_{cs})-\lambda_1({\mathbf{q}}_{sin}))
 \quad \Leftrightarrow \quad
	\frac{\lambda_1({\mathbf{q}}_{sin})+\lambda_2({\mathbf{q}}_{sin})}{2}<\lambda_1({\mathbf{q}}_{cs})
\end{equation}
and scenario (iv) described in Sec.~\ref{gen} is realized. However the last first-order transition (from the EL1 to the SP phase) occurs at small temperature beyond the range of the mean-field theory validity: it follows from Eqs.~\eqref{Fsp} and \eqref{Fel} that ${\cal F}_{sp}$ does not cross ${\cal F}_{el1}$ because
\begin{equation}
  {\cal F}_{sp}-{\cal F}_{el1}=\frac{(\lambda_2({\mathbf{q}}_{sin})-\lambda_1({\mathbf{q}}_{sin}))^2}{8BT}>0,
\end{equation}
where we replace ${\mathbf{q}}_{sp}$ by ${\mathbf{q}}_{sin}$ as it is noted above. The very existence of the transition from the EL1 to the SP phase follows from the fact that the SP state is stable at $T=0$ in the considered ``weak anisotropy'' regime.

We point out that all six scenarios described in Sec.~\ref{gen} can be obtained using a small anisotropic short-range exchange interaction of the form (cf.\ Eq.~\eqref{biax})
\begin{equation}
\label{biax1}
  \mathcal{E}_{an2} = \frac12 \sum_{i,j} \left( E_{ij} \left[ s^x_i s^x_j - s^y_is^y_j\right] - G_{ij} s^z_i s^z_j \right).
\end{equation}
It happens because Fourier components of $E_{ij}$ and $G_{ij}$ become momentum-dependent that enriches the model behavior.

\section{Phase transitions in $\rm \bf MnI_2$}
\label{mni2}

\begin{figure}
  \noindent
  \includegraphics[scale=1]{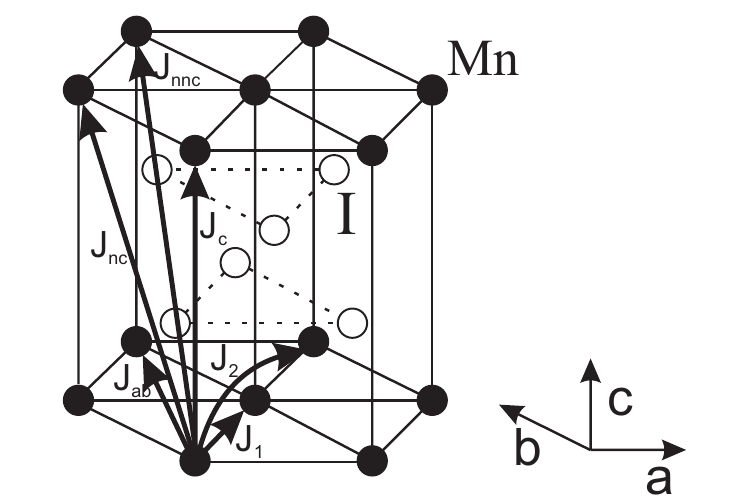}
  \caption{Crystal structure of MnI$_2$. Exchange interactions $J$ are also shown.}
  \label{MnI2}
\end{figure}

MnI$_2$ crystallizes in a hexagonal-layered structure shown in Fig.~\ref{MnI2} with lattice parameters $a=4.146$~\AA\ and $c=6.829$~\AA. \cite{cable} Mn$^{2+}$ ions have spin $S=5/2$ and $g$-factor $g\approx2$. Three successive phase transitions were identified upon temperature decreasing. \cite{sato} At $T_{N1}=3.95$~K, a second-order transition occurs from the paramagnetic state to the incommensurate sinusoidal phase with the modulation vector $\mathbf{q}_{sin}=(0.1025,0.1025,0.5)$. At $T_{N2}=3.8$~K, a second-order transition occurs to another incommensurate sinusoidal phase in which the modulation vector moves continuously from $\mathbf{q}_{sin}$ towards ${\bf q}_{sp}=(0.181,0,0.439)$ upon temperature decreasing. At $T_{N3}=3.45$~K, a jump takes place to a proper screw helical order with the spiral vector ${\bf q}_{sp}$. Spins remain perpendicular to the modulation vectors at $T<T_{N1}$. Then, in the helical phase, spins lie in a plane which is canted from the triangular basal $ab$-plane. One notes that a modified scenario (ii) described in Sec.~\ref{gen} is realized in MnI$_2$ (as compared to scenario (ii), the additional transition arises in MnI$_2$ separating two ICS phases). We demonstrate below that small one-ion anisotropic interactions are responsible for this modification.

\subsection{Basic equations}
\label{eqs}

For the mean-field description of the successive phase transitions in MnI$_2$, we use a model which is based on those proposed before for MnI$_2$ \cite{mni1} and for the isostructural compound $\mathrm{MnBr}_2$ possessing a collinear low-temperature phase rather than the spiral one \cite{mnbr2}. The latter model includes the magnetic dipole interaction, three in-plane exchange interactions and three exchange couplings between spins from neighboring planes (see Fig.~\ref{MnI2}). Notice that interaction $J_{nnc}$ is included because of its straight superexchange path via iodide atoms. This is the only exchange interaction which lowers the sixfold rotational symmetry around the $c$-axis to the threefold one. We take into consideration also small anisotropy terms which are allowed by symmetry: a single-ion easy-axis anisotropy, an in-plane hexagonal anisotropy, and DMI. DMI arises only in the spiral phase (which is ferroelectric in MnI$_2$) due to displacements of iodide atoms removing the inversion symmetry \cite{mni3} (see also Appendix~\ref{dzyal}).

The corresponding mean-field energy reads as
\begin{equation}
\label{ham1}
  \mathcal{E} = \frac12\sum_{i,j} J_{ij} (\mathbf{s}_i \mathbf{s}_j) + \frac12 \sum_{i,j} D^{\alpha \beta}_{ij} s^\alpha_i s^\beta_j - Y \sum_i (s^z_i)^2 - Z \sum_i (s^y_i)^2 \left[ (s^y_i)^2-3(s^x_i)^2 \right]^2+\mathcal{E}_{DM},
\end{equation}
where the first two terms describe the exchange and the dipolar interactions, the third and the fourth terms are the one-ion and the sixfold in-plane anisotropies, respectively, the last term stands for the DMI energy which is discussed below in detail, and a Cartesian coordinate system is implied whose $y$ and $z$ axes coincide with crystallographic $b$ and $c$ ones (see Fig.~\ref{MnI2}), respectively. The characteristic dipolar energy in MnI$_2$ is $\omega_0 \approx 0.31$~K.

Due to the sixfold anisotropy in Eq.~\eqref{ham1}, one has to expand the free energy $\mathcal{F}$ up to the sixth order in $s$ with the result (cf.\ Eq.~\eqref{f1})
\begin{equation}
\label{f}
  \mathcal{F}=\mathcal{E} + A T \sum_i s^2_i + B T \sum_i s^4_i + CT \sum_i s^6_i,
\end{equation}
where $A$, $B$, and $C$ are given by Eqs.~\eqref{a}--\eqref{c}. The Fourier transform of the exchange interaction has the form
\begin{eqnarray}
  J_\mathbf{q}&=&2 \Bigl[J_1(\cos q_a + \cos q_b + \cos(q_a+q_b) ) + J_2 (\cos 2q_a + \cos 2q_b + \cos2(q_a+q_b)) + J_c \cos(q_c) \nonumber \\ \label{Jq}
  &+&  J_{ab} (\cos(2q_a+q_b) + \cos(q_a+2q_b) + \cos(q_a-q_b)) + 2 J_{nc}\cos q_c (\cos q_a + \cos q_b + \cos(q_a+q_b)) \\
  &+& J_{nnc}(\cos(2q_a+q_b-q_c) + \cos(q_a+2q_b+q_c) + \cos(q_a-q_b+q_c))\Bigr]. \nonumber
\end{eqnarray}

\subsection{Sinusoidal phases}

\begin{figure}
  \noindent
  \includegraphics[scale=0.5]{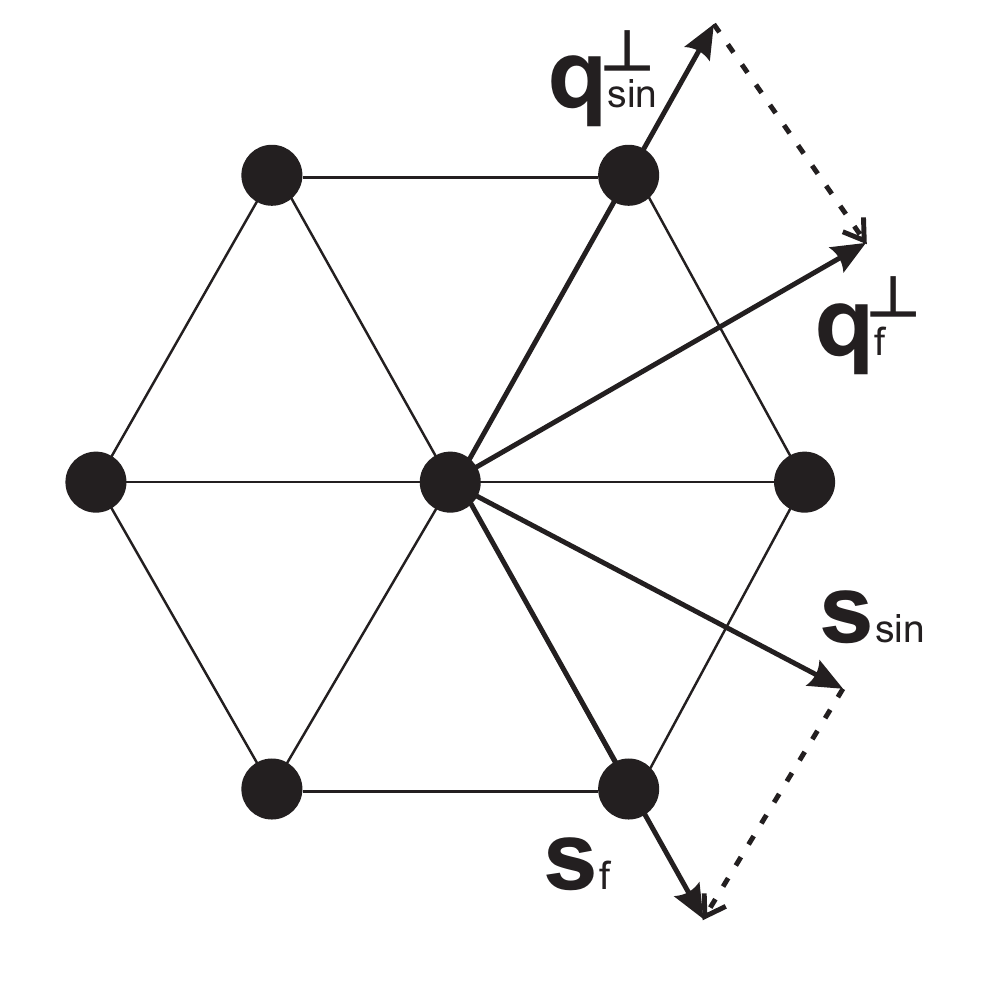}
  \caption{Spin polarization $\bf s$ and the projection ${\bf q}^\perp$ of the modulation vector on the $ab$-plane in two incommensurate sinusoidal phases of $\rm MnI_2$. At $T_{N1}>T>T_{N2}$, ${\bf q}={\bf q}_{sin}$ and ${\bf s}={\bf s}_{sin}$. At $T_{N2}>T>T_{N3}$, $\bf s$ and ${\bf q}^\perp$ rotate continuously upon the temperature decreasing from ${\bf s}_{sin}$ and ${\bf q}_{sin}^\perp$ to ${\bf s}_f$ and ${\bf q}_f^\perp$, respectively. The clockwise rotation presented and the corresponding anticlockwise rotation are equally possible. The spin polarization remains perpendicular to the modulation vector in both incommensurate sinusoidal phases.}
  \label{polar}
\end{figure}

As it is explained above, the transition takes place from the PM phase to the ICS one at $T=T_{N1}$ which is given by Eq.~\eqref{tn1}. Let us consider the spin ordering at $T<T_{N1}$. It depends strongly on the values of the model parameters. However, the range of possible values of the exchange constants is reduced considerably by the requirement that $J_{\bf q}$ should have a minimum at ${\bf q}\approx{\bf q}_{sin}$. Then, we try to reproduce the experimental data by slightly varying the exchange constants and including the small interactions. Our analysis shows that the behavior of $\lambda_1({\mathbf{q}})$ and the corresponding eigenvector are quite simple at a moderate easy-axis anisotropy constant $Y<\omega_0/2$. Dipolar forces make the first eigenvector to be always perpendicular to $\mathbf{q}$ and to lie in the $ab$-plane. This finding is in agreement with experimental data observed in ICS phases \cite{sato}. The difference $\lambda_1({\mathbf{q}})-J_{\bf q}$ is almost independent of $q_z$ and it depends slightly on the value of the $\mathbf{q}$ projection on the $ab$-plane. Then, at temperatures slightly below $T_{N1}$, we obtain a spin texture of the form~\eqref{sdw} with ${\bf q} = {\bf q}_{sin}$ and $\mathbf{a}_1 =s(1,-1,0)$, $\mathbf{a}_2=\bf 0$. The corresponding free energy is given by
\begin{equation}
\label{fics}
  {\cal F}_{ics}^{(1)} = \frac{s^2}{2}(\lambda_1(\mathbf{q}_{sin})+AT)+\frac{3}{8}BT s^4 - \frac{5}{16}Zs^2_y(s^2_y-3s^2_x)^2+\frac{5}{16}CT s^6.
\end{equation}

Notice that the last two terms are negligible in Eq.~\eqref{fics} at $T\approx T_{N1}$ because they are of the sixth order in $s$. However they come into play at lower $T$ upon $s$ growing up. They are indispensable for the description of the experimentally obtained transition at $T=T_{N2}<T_{N1}$ to another ICS phase in which the modulation vector $\bf q$ moves continuously from ${\bf q}_{sin}$ towards ${\bf q}_{sp}$ upon the temperature decreasing at $T_{N2}>T>T_{N3}$. The reason for this moving is simple: the sixfold anisotropy makes directions $[100]$, $[010]$, and $[110]$ to be easy directions for the magnetization. On the other hand, $\bf s$ is directed along the hard $[1\bar 1 0]$ direction at $T\approx T_{N1}$. As a result, the magnetization \eqref{sdw} starts to rotate as it is shown in Fig.~\ref{polar} from $[1\bar 1 0]$ to one of the nearest easy directions ($[ 0 \bar 1 0]$ or $[ 1 0 0]$) at some temperature $T=T_{N2}<T_{N1}$, when the value of the third term in Eq.~\eqref{fics} becomes large enough. The second-order transition at $T=T_{N2}$ is related with the breaking of the two-fold rotational symmetry in the first ICS phase (the magnetization is directed along the twofold symmetry axis of the magnetic subsystem in the first ICS phase, as it is seen from Fig.~\ref{polar}). To demonstrate this, let us consider the correction $\delta{\cal F}$ to free energy \eqref{fics} which arises due to small deviations of $\bf s$ and $\bf q$ from ${\bf s}_{sin}$ and ${\bf q}_{sin}$, respectively,
\begin{equation}
\label{dF}
  \delta {\cal F}
	=
	\frac{c_1}{2} s^2 \delta q^2 - c_2 s \delta s \delta q
	+
	\delta s^2 \left( \frac12 \left( J_{\mathbf{q}_{sin}} - c_3 + AT \right) + \frac{3}{4}BT s^2 + \frac{15}{16}CT s^4 - \frac{45}{16} Z s^4  \right)
	+
	\delta s^4 \frac38\left( BT+\frac52 s^2(2Z+CT)
	\right),
\end{equation}
where ${\bf s} = {\bf s}_{sin} + \delta{\bf s}$, $\delta{\bf s}\perp {\bf s}_{sin}$, $|{\bf s}_{sin}|=s$ is given by Eq.~\eqref{s}, ${\bf q}={\bf q}_{sin}+\delta{\bf q}$, and $c_{1,2,3}$ are some coefficients which are positive in $\rm MnI_2$. Minimization of Eq.~\eqref{dF} with respect of $\delta q$ gives $\delta q = \delta s c_2/(c_1s)$. Substituting the latter equality to Eq.~\eqref{dF}, one finds that the coefficient before $\delta s^2$ becomes negative at $T<T_{N2}$ signifying the second-order transition at $T=T_{N2}$, where
\begin{equation}
\label{tn2}
  T_{N2} \approx T_{N1}\left( 1 + \sqrt{\frac{2B^2 (A T_{N1}-\varkappa)}{5 A^2 Z} }\right)^{-1},
\end{equation}
$\varkappa = J_{\mathbf{q}_{sin}} - c_3-c_2^2/c_1$ and we neglect terms proportional to $C$ which are negligible in $\rm MnI_2$ as specific calculations show. The modulation vector $\bf q$ remains perpendicular to the magnetization in both ICS states in order to minimize the exchange and the dipolar energy.


\subsection{The proper screw spiral phase}

The first-order transition is observed in MnI$_2$ from the second ICS phase to the proper screw spiral phase. The plane in which spins lie in the SP phase does not coincide with the $ab$-plane. The free energy of this phase reads as
\begin{equation}
    \label{fsp}
    {\cal F}_{sp} = s^2 \left[J(\mathbf{q}) + \frac{1}{4} D^{\alpha\beta}_\mathbf{q} v^\alpha_{sp} v^{\beta*}_{sp}-\frac{Y}{2}\sin^2 \theta + AT\right] +BT s^4 - Z s^6 f(\theta,\varphi)+CT s^6 + {\cal E}_{DM},
\end{equation}
where $\theta$ and $\varphi$ are spherical angles determining the normal to the plane in which spins lie,
${\bf v}_{sp} = (\cos \theta \cos \varphi  + i \sin \varphi,
\cos \theta \sin \varphi -i \cos \varphi ,
-\sin \theta )$,
$f(\theta,\varphi) = ( 294 + 171 \cos 2\theta + 42 \cos 4\theta + 5 \cos 6\theta +
 160\cos 6\varphi \sin^6 \theta)/1024$, and the spin ordering of the form \eqref{sdw} is assumed with $\mathbf{a}_1 \perp \mathbf{a}_2 $ and $|\mathbf{a}_1 | = | \mathbf{a}_2|$. The easy-axis anisotropy $Y$ produces the canting of the plane in which spins lie from the $ab$-plane (spins would lie in the $ab$-plane in the spiral phase if $Y$ was zero). In Appendix~\ref{dzyal}, we carry out a phenomenological consideration of DMI in MnI$_2$ based on available experimental data and show that ${\cal E}_{DM}$ in Eq.~\eqref{fsp} has the form
\begin{equation}
\label{Edm}
    {\cal E}_{DM}= - 2s^2 D \sin \left(\frac{\sqrt{3}}{2} q_x \right) \cos\theta.
\end{equation}

\subsection{Results of numerical calculations}
\label{res}

\begin{figure}
  \noindent
  \includegraphics[scale=1]{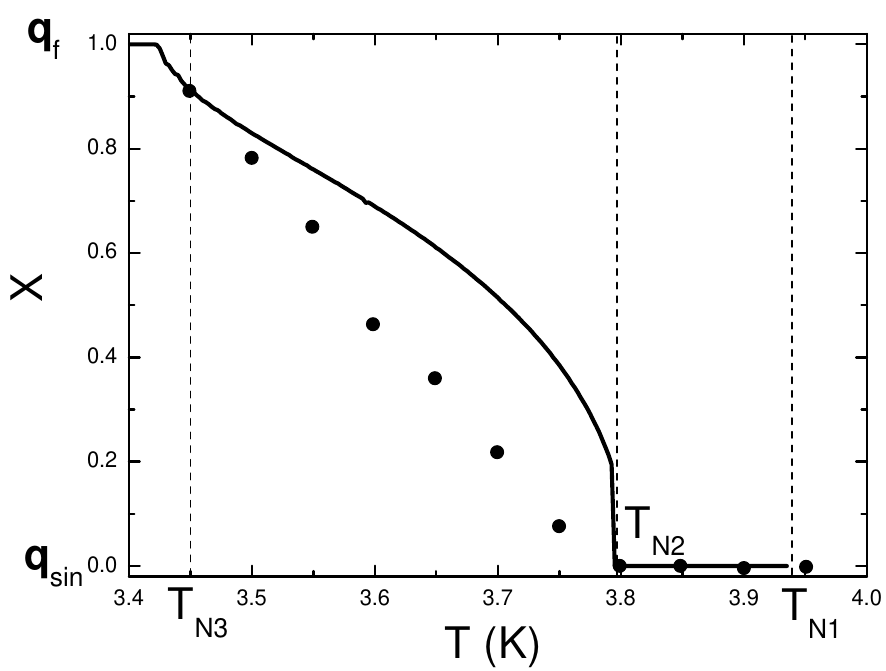}
  \caption{Plot of $X(T)$ which parametrizes the evolution of the modulation vector $\bf q$ upon the temperature decreasing in incommensurate sinusoidal phases of $\rm MnI_2$ (see Eq.~\eqref{q}). At $T_{N3}<T<T_{N2}$, the trajectory of $\bf q$ is almost straight in the reciprocal space which starts at $\mathbf{q}_{sin}=(0.1025,0.1025,0.5)$ and finishes at $\mathbf{q}_f=(0.167,0,0.442)$. Experimental data taken from Ref.~\cite{sato} are shown by circles.}
  \label{traj}
\end{figure}

We obtain the following set of parameters using which the above theory reproduces quantitatively almost all the essential features of phase transitions in MnI$_2$:
\begin{equation}
\label{param1}
\begin{aligned}
  J_1&=-0.13, \quad J_2=0.1, \quad J_{ab}=-0.04, \\
  J_c&=0.04, \quad J_{nc}=-0.0084, \quad J_{nnc}=0.0036, \\
  Y&=0.05, \quad Z=0.015,
\end{aligned}
\end{equation}
where all values are in Kelvins. Eqs.~\eqref{tn1} and \eqref{tn2} reproduce accurately transition temperatures to both ICS phases $T_{N1}=3.95$~K and $T_{N2}=3.8$~K. The trajectory of the modulation vector $\bf q$ in the second ICS phase is almost straight in the reciprocal space. It can be described as
\begin{equation}
\label{q}
	\mathbf{q} \approx (1-X(T))\mathbf{q}_{sin}+X(T)\mathbf{q}_f,
\end{equation}
where $\mathbf{q}_{sin}=(0.1025,0.1025,0.5)$ and $\mathbf{q}_f=(0.167,0,0.442)$ are the initial and the final modulation vectors, correspondingly (see Figs.~\ref{polar} and \ref{traj}). This behavior of $\bf q$ is in a good quantitative agreement with experimental data from Ref.~\cite{sato}. One finds for coefficients in Eqs.~\eqref{dF} and \eqref{tn2}: $c_1 \approx 0.08$~K, $c_2 \approx 0.045$~K, $c_3 \approx 0.007$~K, and $\varkappa \approx 0.62$~K.

Unfortunately, the above formulas failed to describe quantitatively the experimentally observed first-order transition at $T_{N3}\approx3.45$~K to the spiral phase which would take place as a result of the free energies ${\cal F}_{ics}$ and ${\cal F}_{sp}$ crossing. The reason is that our theory is actually based on the expansion in powers of $s/S$ whereas this parameter reaches the value of 0.6 at $T\approx T_{N3}$. We find by minimizing energy \eqref{ham1} at $T=0$ that the following set of parameters gives the proper screw spiral ordering with ${\bf q}_{sp}=(0.166,0,0.428)$ (the latter is very close to the experimentally observed value of $(0.181,0,0.439)$):
\begin{equation}
\label{param2}
\begin{aligned}
  J_1&=-0.105, \quad J_2=0.095, \quad J_{ab}=-0.025, \\
  J_c&=0.06, \quad J_{nc}=-0.0008, \quad J_{nnc}=0.03, \\
  Y&=0.122, \quad Z=0.015, \quad D=0.001,
\end{aligned}
\end{equation}
where all values are in Kelvins. Notice that DMI plays a minor role in the stabilization of the experimentally observed spiral structure at small $T$.

\section{Summary and conclusion}
\label{conc}

To summarize, we discuss within the mean-field theory the impact of the dipolar interaction on critical properties of frustrated Heisenberg spiral antiferromagnets. We demonstrate that dipolar forces turn the single second-order temperature transition from the paramagnetic phase to the spiral one into a sequence of phase transitions of the first and of the second orders. We distinguish six possible scenarios of the successive phase transitions and possible intermediate phases which are summarized in Fig.~\ref{scen}. To the best of our knowledge, at least four of these scenarios were observed before experimentally in specific compounds (e.g., MnBr$_2$, MnI$_2$, $\rm MnWO_4$, and $\rm RbFeCl_3$). We find that not all of these scenarios and intermediate phases can be obtained by replacing the long-range dipolar forces by one-ion anisotropy interactions. In contrast, all the essential features obtained can be reproduced qualitatively by proper short-range exchange anisotropy terms in the Hamiltonian.

We examine using the mean-field theory phase transitions in multiferroic MnI$_2$ showing incommensurate spiral ordering at $T=0$. We reproduce quantitatively the majority of experimental findings observed in this compound. It is shown that the dipolar interaction plays the crucial role in producing the sequence of phase transitions found experimentally. However small symmetry-allowed short-range anisotropic interactions should be also taken into account which lead also to a modification of the corresponding scenario of phase transitions: the additional second-order transition arises separating two different ICS phases.

\begin{acknowledgments}

We thank D.N.~Aristov for discussion. This work is supported by Russian Science Foundation (grant No.\ 14-22-00281).

\end{acknowledgments}

\appendix

\section{Mean-field expansion of the free energy}
\label{mf}

The mean-field expansion of the free energy $\mathcal{F}$ in powers of $s$ can be carried out in Heisenberg antiferromagnets with dipolar forces as it is done, e.g., in Ref.~\cite{gekht2}. The effective mean-field Hamiltonian reads as
\begin{equation}
  \mathcal{H}_{eff}=-\sum_i \mathbf{H}_i \mathbf{S}_i,
\end{equation}
where $\mathbf{H}_i$ is the effective field (see Eq.~\eqref{ham11})
\begin{equation}
  H^\alpha_i = -\frac{1}{2}\sum_{j \beta} D^{\alpha \beta}_{ij} s^\beta_j - \frac{1}{2}\sum_{j} J_{ij} s^\alpha_j.
\end{equation}
One obtains from the partition function
$
  Z = Sp \left( e^{-\mathcal{H}_{eff}/T} \right)
$
for the magnetization at $i$-th site
\begin{equation} \label{mean_s}
  s^\alpha_i=\frac{\partial \ln Z}{\partial H^\alpha_i/T}=\frac{H^\alpha_i}{H_i}S B_S\left( \frac{H_i S}{T}\right),
\end{equation}
where
\begin{equation}
  B_S(x)=\frac{2S+1}{2S} \coth \left(\frac{2S+1}{2S}x\right) - \frac{1}{2S} \coth \left(\frac{x}{2S}\right)
\end{equation}
is the Brillouin function. One infers from Eq.~\eqref{mean_s} that ${\bf s}_i \| {\bf H}_i$,
$
  \partial \mathcal{F}=-\sum_{i \alpha} s^\alpha_i \partial H^\alpha_i,
$
and
\begin{equation}
\label{ff}
  \mathcal{F} = -\sum_{i \alpha} \int^{H^\alpha_i}_0 s^\alpha_i d H^\alpha_i
	= -\sum_{i \alpha} s^\alpha_i H^\alpha_i +  \sum_{i \alpha} \int^{s^\alpha_i}_0 H^\alpha_i d s^\alpha_i
	= \mathcal{E} + T \sum_i \int^{s_i/S}_0 B^{-1}_S(x) d x,
\end{equation}
where $\mathcal{E}$ is given by Eq.~\eqref{ham11}, $B^{-1}_S(x)$ is the inverse of the Brillouin function. Using expansion
\begin{equation}
  B^{-1}_S(x)=\frac{3S}{S+1} x + \frac{9((2S+1)^4-1)}{80 (S+1)^4}x^3 + \frac{9 S^6}{5(S+1)^6}\left[\frac{9S}{5(S+1)}\left(\frac{(2S+1)^4-1}{(2S)^4}\right)^2-\frac{6((2S+1)^6-1)}{7(2S)^6}\right] x^5 + o(x^6),
\end{equation}
one comes from Eq.~\eqref{ff} to Eqs.~\eqref{f1} and \eqref{f}, where
\begin{eqnarray}
\label{a}
  A&=&\frac{3}{2S(S+1)}, \\
\label{b}
  B&=&\frac{9((2S+1)^4-1)}{20(2S)^4(S+1)^4}, \\
\label{c}
  C&=&\frac{3}{10(S+1)^6}\left[\frac{9S}{5(S+1)}\left(\frac{(2S+1)^4-1}{(2S)^4}\right)^2-\frac{6((2S+1)^6-1)}{7(2S)^6}\right].
\end{eqnarray}

\section{Dzyaloshinsky-Moria interaction in the spiral phase of $\rm \bf MnI_2$}
\label{dzyal}

\begin{figure}
  \noindent
  \includegraphics[scale=0.5]{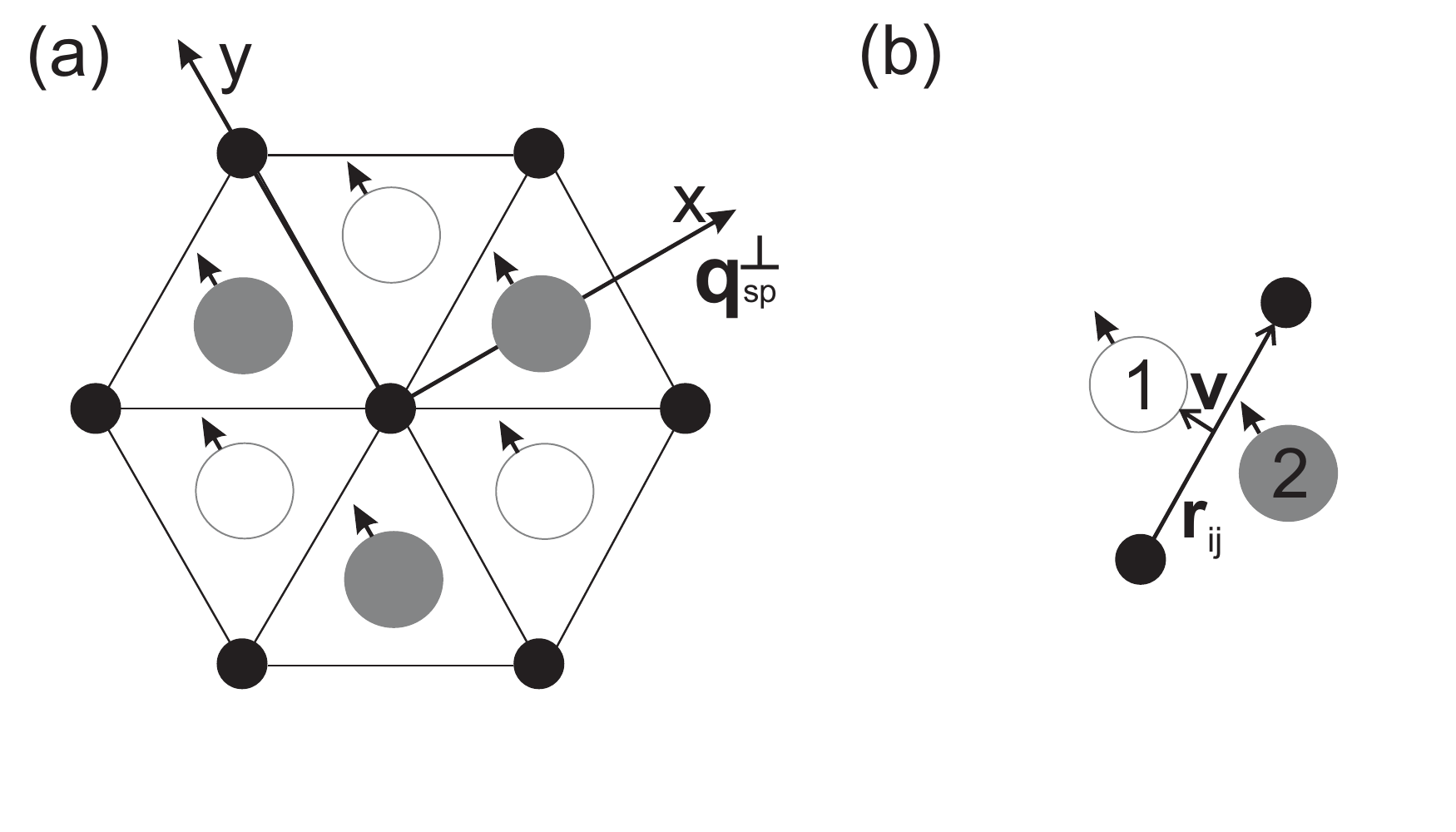}
  \caption{(a) MnI$_2$ in the ferroelectric spiral phase in which the electric polarization is parallel to $y$ axis. The projection of the spiral vector ${\bf q}_{sp}^\perp$ on the $ab$ plane is shown which is directed along the $x$ axis. White and gray big circles are iodide ions which lie above and below the $ab$ plane, correspondingly (see also Fig.~\ref{MnI2}). Shifts of iodide ions are depicted by arrows which are discussed in the text. (b) Illustration of how these shifts break the inversion symmetry between two manganese ions and lead to the Dzyaloshinskii-Moriya interaction. }
  \label{DMI}
\end{figure}

It is obtained experimentally that the phase with the spiral magnetic order is ferroelectric in MnI$_2$. \cite{mni3} The helical magnetic order breaks almost all symmetry elements: as soon as the in-plane projection of $\mathbf{q}_{sp}$ is directed along the $x$ axis, only the twofold rotational symmetry with respect to the $y$ axis remains (see Fig.~\ref{DMI}(a)). This symmetry element allows the electric polarization to be directed along the $y$ axis. This conclusion is in agreement with the experimental observation of Ref.~\cite{mni3}. Bearing in mind also that the spin-orbit coupling leads to the ferroelectricity in MnI$_2$, \cite{mni1} one infers that the inverse Dzyaloshinskii-Moriya mechanism is responsible for the electric polarization. \cite{mni3} A contribution to the polarization from a couple of neighboring spins reads as \cite{nagaosa}
\begin{equation}
  \label{pol}
  \mathbf{p}_{ij}\propto\mathbf{e}_{ij} \times [\mathbf{s}_i \times \mathbf{s}_j],
\end{equation}
where $\mathbf{e}_{ij}=\mathbf{r}_{ij}/r_{ij}$ and $\mathbf{r}_{ij}$ is a vector connecting sites $i$ and $j$. Let us consider the grey iodide ion lying on the $x$ axis shown in Fig.~\ref{DMI}(a) and calculate contributions to the polarization $\bf p$ from three spin pairs adjacent to this iodide ion which are presented in Fig.~\ref{DMI}(a). Using Eq.~\eqref{pol}, we find that one spin pair does not contribute to $\bf p$ because spins are collinear in this pair whereas one has for the rest two spin pairs
\begin{equation}
\label{ss}
  \mathbf{s}_i \times \mathbf{s}_j = s^2 \sin \left(\frac{\sqrt{3}}{2} q_x \right) \mathbf{n},
\end{equation}
where $\mathbf{n}=(\sin\theta\cos\varphi,\sin\theta\sin\varphi,\cos\theta)$ is a unite vector that is normal to the plane in which spins lie. Then, one obtains from Eqs.~\eqref{pol} and \eqref{ss} for the contribution to $\bf p$ related to one iodide ion
\begin{equation}
\label{p}
  \mathbf{p} \propto s^2 \sin \left(\frac{\sqrt{3}}{2} q_x \right) \left[ (\sqrt{3}/2,1/2,0) \times \mathbf{n} + (\sqrt{3}/2,-1/2,0) \times \mathbf{n} \right]
	\propto  \mathbf{e}_ys^2 \sin \left(\frac{\sqrt{3}}{2} q_x \right) \cos\theta,
\end{equation}
where $\mathbf{e}_y$ is a unite vector directed along the $y$ axis. Eq.~\eqref{p} is nonzero for the incommensurate proper screw spin helix whose plane is canted from the $ab$ plane. It can be shown that Eq.~\eqref{p} is valid for all iodide ions. Thus, we obtain that Eq.~\eqref{pol} describes correctly the direction of $\bf p$ observed experimentally in MnI$_2$. Then, due to the $C_2$ symmetry of the $y$ axis and the translational invariance, iodide ions should shift as it is shown in Fig.~\ref{DMI}(a). As DMI vector $\mathbf{D}_{ij}$ in DMI between a pair of spins related with one iodide ion is proportional to $\mathbf{r}_{ij} \times \mathbf{v}$ (see Fig.~\ref{DMI}(b)), impacts to $\mathbf{D}_{ij}$ from two iodide ions (shown in Fig.~\ref{DMI} in grey and white) would cancel each other if there were no these displacements. Then, iodide ions shift produces the electric polarization and $\mathbf{D}_{ij} \propto \mathbf{r}_{ij}\times \mathbf{e}_y$ which is parallel to the $c$ axis. As a result, one comes to Eq.~\eqref{Edm} for ${\cal E}_{DM}$.

\bibliography{Phasebib}

\begin{thebibliography}{19}
\expandafter\ifx\csname natexlab\endcsname\relax\def\natexlab#1{#1}\fi
\expandafter\ifx\csname bibnamefont\endcsname\relax
  \def\bibnamefont#1{#1}\fi
\expandafter\ifx\csname bibfnamefont\endcsname\relax
  \def\bibfnamefont#1{#1}\fi
\expandafter\ifx\csname citenamefont\endcsname\relax
  \def\citenamefont#1{#1}\fi
\expandafter\ifx\csname url\endcsname\relax
  \def\url#1{\texttt{#1}}\fi
\expandafter\ifx\csname urlprefix\endcsname\relax\def\urlprefix{URL }\fi
\providecommand{\bibinfo}[2]{#2}
\providecommand{\eprint}[2][]{\url{#2}}

\bibitem[{\citenamefont{Balents}(2010)}]{balents}
\bibinfo{author}{\bibfnamefont{L.}~\bibnamefont{Balents}},
  \bibinfo{journal}{Nature} \textbf{\bibinfo{volume}{464}},
  \bibinfo{pages}{199} (\bibinfo{year}{2010}).

\bibitem[{\citenamefont{Kawamura}(1998)}]{kawa}
\bibinfo{author}{\bibfnamefont{H.}~\bibnamefont{Kawamura}},
  \bibinfo{journal}{Journal of Physics: Condensed Matter}
  \textbf{\bibinfo{volume}{10}}, \bibinfo{pages}{4707} (\bibinfo{year}{1998}),
  \bibinfo{note}{and references therein}.

\bibitem[{\citenamefont{Sorokin and Syromyatnikov}(2012)}]{sasha}
\bibinfo{author}{\bibfnamefont{A.~O.} \bibnamefont{Sorokin}} \bibnamefont{and}
  \bibinfo{author}{\bibfnamefont{A.~V.} \bibnamefont{Syromyatnikov}},
  \bibinfo{journal}{Phys. Rev. B} \textbf{\bibinfo{volume}{85}},
  \bibinfo{pages}{174404} (\bibinfo{year}{2012}), \bibinfo{note}{and references
  therein}.

\bibitem[{\citenamefont{Shiba}(1982)}]{shiba}
\bibinfo{author}{\bibfnamefont{H.}~\bibnamefont{Shiba}},
  \bibinfo{journal}{Solid State Communications} \textbf{\bibinfo{volume}{41}},
  \bibinfo{pages}{511} (\bibinfo{year}{1982}).

\bibitem[{\citenamefont{Gekht}(1989)}]{gekht}
\bibinfo{author}{\bibfnamefont{R.~S.} \bibnamefont{Gekht}},
  \bibinfo{journal}{Soviet Physics Uspekhi} \textbf{\bibinfo{volume}{32}},
  \bibinfo{pages}{871} (\bibinfo{year}{1989}), \bibinfo{note}{and references
  therein}.

\bibitem[{\citenamefont{Tokura et~al.}(2014)\citenamefont{Tokura, Seki, and
  Nagaosa}}]{nagaosa}
\bibinfo{author}{\bibfnamefont{Y.}~\bibnamefont{Tokura}},
  \bibinfo{author}{\bibfnamefont{S.}~\bibnamefont{Seki}}, \bibnamefont{and}
  \bibinfo{author}{\bibfnamefont{N.}~\bibnamefont{Nagaosa}},
  \bibinfo{journal}{Reports on Progress in Physics}
  \textbf{\bibinfo{volume}{77}}, \bibinfo{pages}{076501}
  (\bibinfo{year}{2014}), \bibinfo{note}{and references therein}.

\bibitem[{\citenamefont{Sato et~al.}(1995)\citenamefont{Sato, Kadowaki, and
  Iio}}]{sato}
\bibinfo{author}{\bibfnamefont{T.}~\bibnamefont{Sato}},
  \bibinfo{author}{\bibfnamefont{H.}~\bibnamefont{Kadowaki}}, \bibnamefont{and}
  \bibinfo{author}{\bibfnamefont{K.}~\bibnamefont{Iio}},
  \bibinfo{journal}{Physica B: Condensed Matter}
  \textbf{\bibinfo{volume}{213}}, \bibinfo{pages}{224 } (\bibinfo{year}{1995}).

\bibitem[{\citenamefont{Cable et~al.}(1962)\citenamefont{Cable, Wilkinson,
  Wollan, and Koehler}}]{cable}
\bibinfo{author}{\bibfnamefont{J.~W.} \bibnamefont{Cable}},
  \bibinfo{author}{\bibfnamefont{M.~K.} \bibnamefont{Wilkinson}},
  \bibinfo{author}{\bibfnamefont{E.~O.} \bibnamefont{Wollan}},
  \bibnamefont{and} \bibinfo{author}{\bibfnamefont{W.~C.}
  \bibnamefont{Koehler}}, \bibinfo{journal}{Phys. Rev.}
  \textbf{\bibinfo{volume}{125}}, \bibinfo{pages}{1860} (\bibinfo{year}{1962}).

\bibitem[{\citenamefont{Wu et~al.}(2012)\citenamefont{Wu, Cai, Xie, Weng, Fan,
  and Hu}}]{mni1}
\bibinfo{author}{\bibfnamefont{X.}~\bibnamefont{Wu}},
  \bibinfo{author}{\bibfnamefont{Y.}~\bibnamefont{Cai}},
  \bibinfo{author}{\bibfnamefont{Q.}~\bibnamefont{Xie}},
  \bibinfo{author}{\bibfnamefont{H.}~\bibnamefont{Weng}},
  \bibinfo{author}{\bibfnamefont{H.}~\bibnamefont{Fan}}, \bibnamefont{and}
  \bibinfo{author}{\bibfnamefont{J.}~\bibnamefont{Hu}}, \bibinfo{journal}{Phys.
  Rev. B} \textbf{\bibinfo{volume}{86}}, \bibinfo{pages}{134413}
  (\bibinfo{year}{2012}).

\bibitem[{\citenamefont{Xiang et~al.}(2011)\citenamefont{Xiang, Kan, Zhang,
  Whangbo, and Gong}}]{mni2}
\bibinfo{author}{\bibfnamefont{H.~J.} \bibnamefont{Xiang}},
  \bibinfo{author}{\bibfnamefont{E.~J.} \bibnamefont{Kan}},
  \bibinfo{author}{\bibfnamefont{Y.}~\bibnamefont{Zhang}},
  \bibinfo{author}{\bibfnamefont{M.-H.} \bibnamefont{Whangbo}},
  \bibnamefont{and} \bibinfo{author}{\bibfnamefont{X.~G.} \bibnamefont{Gong}},
  \bibinfo{journal}{Phys. Rev. Lett.} \textbf{\bibinfo{volume}{107}},
  \bibinfo{pages}{157202} (\bibinfo{year}{2011}).

\bibitem[{\citenamefont{Kurumaji et~al.}(2011)\citenamefont{Kurumaji, Seki,
  Ishiwata, Murakawa, Tokunaga, Kaneko, and Tokura}}]{mni3}
\bibinfo{author}{\bibfnamefont{T.}~\bibnamefont{Kurumaji}},
  \bibinfo{author}{\bibfnamefont{S.}~\bibnamefont{Seki}},
  \bibinfo{author}{\bibfnamefont{S.}~\bibnamefont{Ishiwata}},
  \bibinfo{author}{\bibfnamefont{H.}~\bibnamefont{Murakawa}},
  \bibinfo{author}{\bibfnamefont{Y.}~\bibnamefont{Tokunaga}},
  \bibinfo{author}{\bibfnamefont{Y.}~\bibnamefont{Kaneko}}, \bibnamefont{and}
  \bibinfo{author}{\bibfnamefont{Y.}~\bibnamefont{Tokura}},
  \bibinfo{journal}{Phys. Rev. Lett.} \textbf{\bibinfo{volume}{106}},
  \bibinfo{pages}{167206} (\bibinfo{year}{2011}).

\bibitem[{\citenamefont{Lautenschl\"ager
  et~al.}(1993{\natexlab{a}})\citenamefont{Lautenschl\"ager, Weitzel, Vogt,
  Hock, B\"ohm, Bonnet, and Fuess}}]{mnw1}
\bibinfo{author}{\bibfnamefont{G.}~\bibnamefont{Lautenschl\"ager}},
  \bibinfo{author}{\bibfnamefont{H.}~\bibnamefont{Weitzel}},
  \bibinfo{author}{\bibfnamefont{T.}~\bibnamefont{Vogt}},
  \bibinfo{author}{\bibfnamefont{R.}~\bibnamefont{Hock}},
  \bibinfo{author}{\bibfnamefont{A.}~\bibnamefont{B\"ohm}},
  \bibinfo{author}{\bibfnamefont{M.}~\bibnamefont{Bonnet}}, \bibnamefont{and}
  \bibinfo{author}{\bibfnamefont{H.}~\bibnamefont{Fuess}},
  \bibinfo{journal}{Phys. Rev. B} \textbf{\bibinfo{volume}{48}},
  \bibinfo{pages}{6087} (\bibinfo{year}{1993}{\natexlab{a}}).

\bibitem[{\citenamefont{Arkenbout et~al.}(2006)\citenamefont{Arkenbout,
  Palstra, Siegrist, and Kimura}}]{mnw2}
\bibinfo{author}{\bibfnamefont{A.~H.} \bibnamefont{Arkenbout}},
  \bibinfo{author}{\bibfnamefont{T.~T.~M.} \bibnamefont{Palstra}},
  \bibinfo{author}{\bibfnamefont{T.}~\bibnamefont{Siegrist}}, \bibnamefont{and}
  \bibinfo{author}{\bibfnamefont{T.}~\bibnamefont{Kimura}},
  \bibinfo{journal}{Phys. Rev. B} \textbf{\bibinfo{volume}{74}},
  \bibinfo{pages}{184431} (\bibinfo{year}{2006}).

\bibitem[{\citenamefont{Heyer et~al.}(2006)\citenamefont{Heyer, Hollmann,
  Klassen, Jodlauk, Bohatý, Becker, Mydosh, Lorenz, and Khomskii}}]{mnw3}
\bibinfo{author}{\bibfnamefont{O.}~\bibnamefont{Heyer}},
  \bibinfo{author}{\bibfnamefont{N.}~\bibnamefont{Hollmann}},
  \bibinfo{author}{\bibfnamefont{I.}~\bibnamefont{Klassen}},
  \bibinfo{author}{\bibfnamefont{S.}~\bibnamefont{Jodlauk}},
  \bibinfo{author}{\bibfnamefont{L.}~\bibnamefont{Bohatý}},
  \bibinfo{author}{\bibfnamefont{P.}~\bibnamefont{Becker}},
  \bibinfo{author}{\bibfnamefont{J.~A.} \bibnamefont{Mydosh}},
  \bibinfo{author}{\bibfnamefont{T.}~\bibnamefont{Lorenz}}, \bibnamefont{and}
  \bibinfo{author}{\bibfnamefont{D.}~\bibnamefont{Khomskii}},
  \bibinfo{journal}{Journal of Physics: Condensed Matter}
  \textbf{\bibinfo{volume}{18}}, \bibinfo{pages}{L471} (\bibinfo{year}{2006}).

\bibitem[{\citenamefont{Cohen and Keffer}(1955)}]{cohen}
\bibinfo{author}{\bibfnamefont{M.~H.} \bibnamefont{Cohen}} \bibnamefont{and}
  \bibinfo{author}{\bibfnamefont{F.}~\bibnamefont{Keffer}},
  \bibinfo{journal}{Phys. Rev.} \textbf{\bibinfo{volume}{99}},
  \bibinfo{pages}{1128} (\bibinfo{year}{1955}), \bibinfo{note}{and references
  therein}.

\bibitem[{\citenamefont{Sato et~al.}(1994)\citenamefont{Sato, Kadowaki, Masudo,
  and Iio}}]{mnbr2}
\bibinfo{author}{\bibfnamefont{T.}~\bibnamefont{Sato}},
  \bibinfo{author}{\bibfnamefont{H.}~\bibnamefont{Kadowaki}},
  \bibinfo{author}{\bibfnamefont{H.}~\bibnamefont{Masudo}}, \bibnamefont{and}
  \bibinfo{author}{\bibfnamefont{K.}~\bibnamefont{Iio}}, \bibinfo{journal}{J.
  Phys. Soc. Japan} \textbf{\bibinfo{volume}{63}}, \bibinfo{pages}{4583}
  (\bibinfo{year}{1994}).

\bibitem[{\citenamefont{Lautenschl\"ager
  et~al.}(1993{\natexlab{b}})\citenamefont{Lautenschl\"ager, Weitzel, Vogt,
  Hock, B\"ohm, Bonnet, and Fuess}}]{lauten}
\bibinfo{author}{\bibfnamefont{G.}~\bibnamefont{Lautenschl\"ager}},
  \bibinfo{author}{\bibfnamefont{H.}~\bibnamefont{Weitzel}},
  \bibinfo{author}{\bibfnamefont{T.}~\bibnamefont{Vogt}},
  \bibinfo{author}{\bibfnamefont{R.}~\bibnamefont{Hock}},
  \bibinfo{author}{\bibfnamefont{A.}~\bibnamefont{B\"ohm}},
  \bibinfo{author}{\bibfnamefont{M.}~\bibnamefont{Bonnet}}, \bibnamefont{and}
  \bibinfo{author}{\bibfnamefont{H.}~\bibnamefont{Fuess}},
  \bibinfo{journal}{Phys. Rev. B} \textbf{\bibinfo{volume}{48}},
  \bibinfo{pages}{6087} (\bibinfo{year}{1993}{\natexlab{b}}).

\bibitem[{\citenamefont{Gvozdikova et~al.}(2016)\citenamefont{Gvozdikova,
  Ziman, and Zhitomirsky}}]{zh}
\bibinfo{author}{\bibfnamefont{M.~V.} \bibnamefont{Gvozdikova}},
  \bibinfo{author}{\bibfnamefont{T.}~\bibnamefont{Ziman}}, \bibnamefont{and}
  \bibinfo{author}{\bibfnamefont{M.~E.} \bibnamefont{Zhitomirsky}},
  \bibinfo{journal}{Phys. Rev. B} \textbf{\bibinfo{volume}{94}},
  \bibinfo{pages}{020406} (\bibinfo{year}{2016}).

\bibitem[{\citenamefont{Gekht}(1984)}]{gekht2}
\bibinfo{author}{\bibfnamefont{R.~S.} \bibnamefont{Gekht}},
  \bibinfo{journal}{Zh. Eksp. Teor. Fiz.} \textbf{\bibinfo{volume}{87}},
  \bibinfo{pages}{2095} (\bibinfo{year}{1984}).

\end{thebibliography}

\end{document}